\documentclass[prb,twocolumn,showpacs,superscriptaddress]{revtex4-1}
\usepackage{epsfig}
\usepackage{amsmath,amssymb}
\usepackage{graphicx}

\begin{document}

\title{Interplay of Coulomb interaction and spin-orbit effects in multi-level quantum dots}
\author{S. Grap,$^1$ V. Meden,$^1$ and S. Andergassen$^{1,2}$\\
  {\small\em $^1$Institut f\"ur Theorie der Statistischen Physik, RWTH Aachen, 
    D-52056 Aachen, Germany} \\
  {\small\em and JARA-Fundamentals of Future Information Technology} \\
{\small\em $^2$Faculty of Physics, University of Vienna, Boltzmanngasse 5, 1090 Wien, Austria} 
}
\date{\small\today}
\begin{abstract}
We study electron transport through a multi-level quantum dot with Rashba 
spin-orbit interaction in the presence of local Coulomb repulsion. We focus on the parameter regime in which the level spacing is larger than the level broadening.
Motivated by recent experiments, we compute the level splitting induced by the spin-orbit interaction
at finite Zeeman fields $B$, which provides a measure of the renormalized spin-orbit energy. 
This level splitting is responsible for the suppression of the Kondo ridges at finite $B$ 
characteristic for the multi-level structure. In addition, the dependence of renormalized $g$-factors on the relative orientation of the applied $B$ field and the
spin-orbit direction following two different protocols used in experiments is investigated.

\pacs{05.60.Gg, 71.10.-w, 71.70.Ej, 73.63.Kv}
\end{abstract}
\maketitle
\section{Introduction}

In linear response transport through quantum dots at low temperatures, 
a two-fold Kramers degeneracy 
leads to the spin Kondo effect in presence of a sufficiently strong local interaction $U$
as compared to the level-lead hybridization $\Gamma$. In the Kondo regime charge fluctuations of the dot are suppressed and the physics is dominated by spin fluctuations. Varying the level positions by an external gate voltage $V_{\rm G}$, characteristic 
conductance plateaus, so-called Kondo ridges, of width $U$ appear around odd (average) electron 
fillings.\cite{Hewson,Glazman,Ng,Goldhaber,Cronenwett,Schmid,Wiel} 
Breaking the two-fold Kramers-degeneracy 
by a local Zeeman field of amplitude $B$ destroys the Kondo ridge and the conductance plateau is split into two Lorentzian resonances of width $\propto \Gamma$ along the $V_{\rm G}$-axis. 
In contrast, spin-orbit interaction (SOI), 
although breaking spin-rotational symmetry by designating a certain (spin) direction, 
does not destroy the Kondo effect.\cite{DMK,Meir,thesisB,jens} In the presence of SOI spin is no longer 
a good quantum number but a Kramers doublet remains as time-reversal symmetry is 
conserved. 

In multi-level dots with initially (at $B=0$)
well separated levels increasing $B$ might lead to energetically degenerate states 
(level crossings) resulting from different orbitals. If one is a spin-up and 
one a spin-down state and the gate voltage is tuned such that an electron fluctuates 
between these states one might expect the emergence of a spin Kondo effect at finite 
magnetic fields.\cite{DMK,Pustilnik,2,Izumida}
If the orbital quantum number 
is conserved in the leads in such systems additional orbital Kondo 
effects\cite{Cox} and combinations of spin and orbital Kondo effects\cite{Borda} may appear. 
Here we consider a setup where the dot orbital quantum number does not arise in the leads 
and we thus concentrate on the spin Kondo effect.
In contrast to the standard $B=0$ Kondo effect the one appearing at finite $B$ is not protected 
by time-reversal symmetry and can be suppressed in presence of a finite SOI.\cite{grap}

\begin{figure}[t!]
\center{\includegraphics[clip=true,width=80mm]{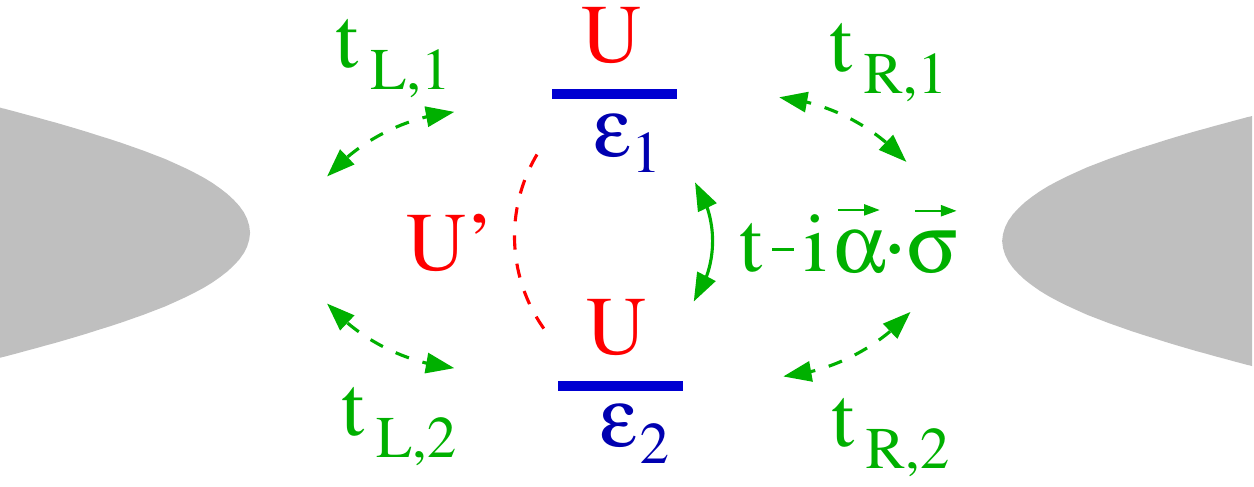}}
\caption{\label{fig:fig1} (Color online) 
  The considered setup consists of two parallel quantum dots with
  energies $\epsilon_{1/2}=V_{\rm G} \pm \delta$ coupled by a hopping
  amplitude $t$ and a SOI of strength $\alpha$. The levels are split by an external Zeeman field $B$.
  The local Coulomb
  interaction is $U$ and the interaction between electrons on the two dots is $U'$.
  The system is coupled to noninteracting leads by hopping amplitudes $t_{\beta,j}$ with $\beta=L,R$ and $j=1,2$.}
\end{figure}

We here study the dot setup sketched in Fig.~\ref{fig:fig1} as a minimal model for a multi-level system. It consists of a tight-binding model with two lattice sites $1$ and $2$ coupled by the
electron hopping of amplitude $t$ and connected to two semi-infinite 
noninteracting leads via tunnel couplings
of strength $t_{\beta,j}$ with $\beta=L,R$ and $j=1,2$.
The on-site energies of the two levels 
are given by $\epsilon_{1/2}=V_{\rm G} \pm \delta$.    
The Rashba SOI identifies the $z$-direction of the spin space and 
is modeled as an imaginary electron hopping with spin-dependent sign between the 
two lattice sites.\cite{Winkler,Mireles,BM1,BM2,thesisB} We choose the parameters in such a way as to deal with two well separated sets of spin degenerate states, that is the molecular regime.
We here exclusively consider the coupling of a magnetic field to the spin 
degree of freedom (Zeeman term) and neglect its effect on the orbital motion.
The Zeeman field can be decomposed in a parallel and an orthogonal component with respect to the $z$-direction.
The local Coulomb interaction (charging energy) is modeled as an on-site $U$ as well as a nearest-neighbor interaction $U'$, 
and treated within an approximate static functional-renormalization group (fRG) 
approach.\cite{KEM} 

Motivated by recent experiments on InAs quantum dots in the Kondo 
regime,\cite{kretinin,kanai,deacon} we determine the 
level splitting induced by the SOI at finite $B$ fields. It provides a measure for the renormalized SOI energy.
The dependence of this spin-orbit energy on the relative orientation of the Zeeman field and the spin-orbit direction was measured.\cite{takahashi,schroer,defranceschi,witek,katsaros,kanai,deacon,Nadj} In the interpretation of the data Kondo correlations were ignored. We here study how those modify the angular dependence of the level splitting. In addition, we investigate how the two-particle interaction affects the orientation dependence of effective $g$-factors, as extracted experimentally following two distinct protocols. In the first the gate-voltage dependence of the linear conductance\cite{nilsson,kanai,deacon}  is used to extract $g_{\rm Cond}$, in the second bias spectroscopy is employed leading to $g_{\rm Levels}$.\cite{csonka,takahashi,defranceschi,kanai}
We will show that the SOI energy and $g_{\rm Levels}$ both have an overall amplitude which is renormalized by the two-particle interaction. In addition, the functional dependence on the angle between the SOI and Zeeman field direction is modified. The $g$-factor $g_{\rm Cond}$ remains unaffected. The renormalization due to the two-particle interaction is seen to be competing with asymmetry effects.

The paper is organized as follows. In the next section, we introduce our minimal multi-level dot 
model and review the basic concepts of the approximate fRG treatment of the Coulomb 
interaction. In Sec.~\ref{sec:res} we discuss our results. 
We first assess the potential 
of our model in connection with the fRG to describe the experimentally observed effects 
in Sec.~\ref{sec:Cond}. In Sec.~\ref{sec:ESOI} we determine the angular 
dependence of the SOI-induced level splitting at finite Zeeman field for the simplest possible model and provide
an intuitive physical picture for the interpretation of the finite-bias spectroscopy. 
In Sec.~\ref{sec:g} we discuss the extraction of effective $g$-factors from the
gate-voltage dependence of the linear conductance and the bias spectroscopy as well as their respective angular dependencies. Sec.~\ref{sec:Asym} deals with an extension of the simplest models and examines a more realistic asymmetric set of parameters, making close contact to recent experimental data.\cite{kanai,takahashi,deacon,defranceschi}
Finally, we conclude with a short summary.

\section{Model and Method}\label{sec:model&method}

\subsection{Multi-Level Quantum Dot}\label{sec:model}

The considered minimal multi-level quantum dot model is realized by two spin-degenerate levels (at $B=0$) with the possibility of electron hoppings between these levels as sketched in Fig.~\ref{fig:fig1}. The Hamiltonian of the isolated dot contains several terms
\begin{equation*}
H_{\rm dot}=H_0+H_{\rm SOI}+H_Z+H_{\rm int} \;.
\end{equation*}
The free part 
\begin{equation*}
H_0=\sum_{\sigma} \left[\sum_{j=1,2} \epsilon_j d_{j,\sigma}^\dagger
  d_{j,\sigma}-t\left(
d_{2,\sigma}^\dagger d_{1,\sigma} +\mbox{H.c.}\right)\right] \; ,
\end{equation*}
with $d_{j,\sigma}^\dagger$ being the creation operator of an electron of spin $\sigma=\uparrow,\downarrow$
on the dot site $j=1,2$ (Wannier states), contains the 
conventional hopping $t>0$, and the on-site energies $\epsilon_{1/2}=V_{\rm G} \pm \delta$
which can be tuned by an external gate voltage $V_{\rm G}$. 
The difference of the on-site energies is parametrized 
by the level splitting $2\delta$. The effect of a Rashba SOI resulting from spatial confinement
is taken into account by an imaginary hopping amplitude of spin-dependent 
sign.\cite{Winkler,Mireles,BM1,BM2,thesisB}
The Rashba hopping term with amplitude $\alpha>0$ reads
\begin{eqnarray}
\label{HSOI}
H_{\rm SOI}&=&\alpha\sum_{\sigma,\sigma'}\left[ d_{2,\sigma}^\dagger
\left(i\sigma_z\right)_{\sigma,\sigma'} d_{1,\sigma'} + \mbox{H.c.} \right] \; ,
\end{eqnarray}
with the third Pauli matrix $\sigma_z$. This choice corresponds to a confinement in $y$-direction if one starts from a one-dimensional system in $x$-direction.\cite{thesisB} We note in passing that other SOI terms similar to $H_{\rm SOI}$ with $i\sigma_z \rightarrow i\sigma_y$ (Rashba SOI from confinement in $z$-direction) or $i\sigma_z \rightarrow \sigma_y$ (Dresselhaus SOI)\cite{Winkler} can be included in the model but will be omitted for simplicity. 
The SOI breaks the spin-rotational invariance and the Zeeman field can be decomposed in a component parallel to the SOI (that is in 
$z$-direction) and one perpendicular to it.\cite{note} We here choose the $x$-direction such that the (local) Zeeman term reads
\begin{eqnarray}
\label{HZ}
H_Z & = &   B \sum_{\sigma,\sigma'} \sum_{j=1,2} \left[  
d_{j,\sigma}^\dag (\sigma_z)_{\sigma,\sigma'} d_{j,\sigma'} {\rm sin} \, \phi  \right. \nonumber 
\\ && \qquad\qquad\;+ \left. 
d_{j,\sigma}^\dag (\sigma_x)_{\sigma,\sigma'} d_{j,\sigma'} {\rm cos} \, \phi \right] \; .
\end{eqnarray}
For $\phi = \pm \pi/2$ the SOI and the $B$-field are (anti-)parallel. In this case physical quantities are similar to the $\alpha=0$ case\cite{thesisB,grap} if $t$ is replaced by the effective hopping $t_{\rm eff}=\sqrt{t^2+\alpha^2}$.
In particular, the Kondo ridges at finite $B$ are preserved (see Sec.~\ref{sec:Cond} for a more detailed discussion).
The local Coulomb interaction is included by
\begin{eqnarray*}
H_{\rm int}&=&U\sum_{j=1,2} \left(n_{j,\uparrow} - \frac{1}{2} \right) \left( n_{j,\downarrow} - \frac{1}{2} \right) 
\nonumber \\ && +\;U' \left( n_{1} -1 \right) \left(n_{2}-1\right) \; ,
\end{eqnarray*}
for the local $U>0$ and nearest-neighbor $U'>0$ interactions respectively, with
$n_{j,\sigma}= d_{j,\sigma}^\dagger  d_{j,\sigma}$ and $n_j= \sum_\sigma n_{j,\sigma}$. In principle different local interactions $U_1\, ,U_2$ on the two sites can be included but we focus on the case $U=U_1=U_2$.
By subtracting $1/2$ from $n_{j,\sigma}$ in the definition of $H_{\rm int}$ the point $V_{\rm G}=0$ corresponds to half-filling (even in the presence of Coulomb repulsion) of a symmetric serial dot, which will be the main geometry under consideration in the following.

Finally, the dot Hamiltonian is supplemented by a term describing two semi-infinite 
noninteracting leads, which we model as one-dimensional 
tight-binding chains 
\begin{eqnarray}
\label{lead}
H_\mathrm{lead} =- \tau \sum_{\beta=L,R}  
\sum_{j=1}^\infty \sum_{\sigma} \left[ c_{\beta,j+1,\sigma}^\dagger 
c_{\beta,j,\sigma} +\mathrm{H.c.} \right]\;,
\end{eqnarray}
with lead operators $ c_{\beta,j,\sigma}^{(\dagger)}$ and equal band width 
$4 \tau$. In the following we choose $\tau=1$ as the energy unit. We note in passing that it is possible to include Rashba SOI terms in the leads. For our case of vanishing magnetic field in the leads the effects of such terms enter our calculation only in terms of an effective hopping.\cite{thesisB} The consequences for
the Kondo temperature were discussed in Refs.~[\onlinecite{Meir}], [\onlinecite{ZitkoBonca}] and [\onlinecite{Zarea}].
The dot-lead couplings are given by the tunnel Hamiltonian
\begin{equation}
\label{Hcoup}
H_\mathrm{coup}=\sum_{\beta=L,R} \sum_{j=1,2}\sum\limits_{\sigma} \left[ t_{\beta,j} \; d_{j,\sigma}^\dagger c_{\beta,1,\sigma}   +\mathrm{H.c.}\right]\;,
\end{equation}
with tunnel barriers set by $t_{\beta,j}$. For simplicity we here consider only real $t_{\beta,j}$\cite{FootnoteCouplings} in the so-called wide-band limit (see e.g. Ref.~[\onlinecite{KEM}]) in which the tunnel barriers only enter in combination 
with the local lead density of states evaluated at the chemical potential. For our setup we find $\Gamma_{\beta,j}= \pi t_{\beta,j}^2\rho_{\rm leads}\stackrel{\mathrm{\tau=1}}=t_{\beta,j}^2$.

\subsection{Functional RG}\label{sec:RG}

We briefly review the applied approximation scheme which is based on the fRG.
\cite{review} Recent applications to systems with SOI include homogeneous quantum wires 
\cite{BM1,BM2} and quantum dots.\cite{grap,preprint}

Starting point of the fRG scheme is the bare ($U=U'=0$) 
propagator ${\mathcal{G}}_0$ 
of the double dot. The leads are projected onto the dot sites and enter via the 
hybridizations $\Gamma_{j}=\sum_{\beta} \Gamma_{\beta, j}$ and $\gamma=\pi \rho_{\rm leads}\sum_{\beta} t_{\beta, 1}t_{\beta, 2}\stackrel{\mathrm{\tau=1}}=\sum_{\beta} t_{\beta, 1}t_{\beta, 2}$.
\cite{KEM} In the basis 
\begin{eqnarray}
\label{basis}
\left\{ \left|1,\uparrow\right>, 
\left|1,\downarrow\right>, \left|2,\uparrow\right>, 
\left|2,\downarrow\right> \right\}
\end{eqnarray}
of single-particle dot states the inverse of the propagator in Matsubara frequency space reads  
\begin{widetext}
 \begin{eqnarray}
&&\!\!\!\!\!\!\!\! {\mathcal{G}}_0^{-1}(i\omega) = \nonumber  \\ && \left( \begin{array}{cccc}
 i\omega-\epsilon_1-B \sin \phi + i \Gamma_1(\omega)\!\!\!\!& -B \cos \phi &t-i\alpha+i \gamma(\omega)&0 \\
 - B \cos\phi&i\omega-\epsilon_1+ B \sin \phi + i \Gamma_1(\omega)\!\!\!\! &0&t+i\alpha +i \gamma(\omega)\\
 t+i\alpha+i \gamma(\omega)&0&i\omega-\epsilon_2-B \sin \phi+i \Gamma_2(\omega)\!\!\!\!&- B \cos \phi\\
0& t-i\alpha+i \gamma(\omega) &-B{\rm cos} \, \phi& i\omega-\epsilon_2+B \sin \phi+i 
\Gamma_2(\omega)\end{array} \right) \label{g0}
\ ,\ \; 
\end{eqnarray}
\end{widetext}
with $\Gamma_{j}(\omega)= \Gamma_{j} \, \mbox{sgn}(\omega)$ and 
$ \gamma(\omega)=\gamma\, \mbox{sgn}(\omega)$.
Within the fRG ${\mathcal{G}}_0$ is replaced by a modified propagator ${\mathcal{G}}_0^\Lambda$ which suppresses low-energy degrees of freedom below a sharp Matsubara frequency cutoff $\Lambda$:
\begin{equation*}
\mathcal{G}_0^\Lambda(i\omega)=\Theta(|\omega|-\Lambda)\mathcal{G}_0(i\omega)\;.
\end{equation*}
The cutoff $\Lambda$ is sent from $\infty$ down to $0$, at which the cutoff-free problem is restored. 
Inserting $\mathcal{G}_0^\Lambda$ in the generating functional of the one-particle irreducible vertex 
functions, an infinite hierarchy of coupled differential equations is obtained by differentiating 
the generating functional with respect to $\Lambda$ and expanding it in powers of the external fields. 
Practical implementations require a truncation of the flow equation hierarchy.

Following Ref.~[\onlinecite{grap}], 
we restrict the present analysis to the first order in the hierarchy and only consider the flow of the
single-particle vertex, that is the self-energy $\Sigma^{\Lambda}$. Within this truncation $\Sigma^{\Lambda}$ is frequency independent leading to a static approximation. It already captures the relevant Kondo physics present in the system and allows for a qualitative description of equilibrium properties such as the linear conductance or the dot occupation with minor numerical effort. In the case of a single (spin-degenerate) dot level comparing with
numerical renormalization group data and Bethe ansatz
results\cite{KEM} shows, that the fRG is reliable if
$U/\Gamma$ (with $\Gamma=\Gamma_L + \Gamma_R$) is not too large. In
particular, the results for the zero-temperature linear conductance and
the renormalized effective single-particle level show Kondo physics
(Kondo ridges and pinning of the level energy at the Fermi energy). The
Zeeman field necessary to suppress the conductance at half filling to
half its value can be used to define a Kondo scale $T_{\rm K}$. Within
first order fRG it is given by $T_{\rm K} \sim \exp{[-U/(\pi\Gamma)]}$
compared to the exact result\cite{Hewson}  $T_{\rm K} \sim \exp{[-\pi
U/(8\Gamma)]}$. First order fRG was also shown to produce reliable
results for multi-level dots as long as the same constraint on the ratio
of the local Coulomb interaction and the hybridization as above is
fulfilled and the number of degenerate single-particle levels does not
become too large.\cite{KEM}
The quantitative accuracy can be improved by including the flow of the static 
part of the two-particle vertex (effective interaction),\cite{AEM,KEM,preprint} which is beyond the 
scope of the present analysis. For an in depth discussion concerning the range of validity of these approximations see Ref.~[\onlinecite{KEM}].

The flow equation for the self-energy reads\cite{KEM} 
\begin{equation}
\frac{\partial}{\partial\Lambda}\Sigma_{a',a}^\Lambda=-\frac{1}{2\pi}\sum\limits_{\omega=\pm\Lambda}\sum\limits_{b,b'}e^{i\omega 0^+}{\mathcal{G}}_{b,b'}^\Lambda(i\omega)\Gamma_{a',b';a,b}  \label{DGLsigma}
\end{equation} 
where the indices $a,a',b,b'$ label the quantum numbers $({j,\sigma})$. $\Gamma_{a',b';a,b}$ is
the anti-symmetrized two-particle vertex, and 
the interacting Green function ${\mathcal{G}}$ is determined by the Dyson equation
\begin{equation}
\label{dyson}
{\mathcal{G}}^\Lambda(i\omega)=\left[\mathcal{G}_0^{-1}(i\omega)-\Sigma^\Lambda\right]^{-1} \; .
\end{equation}
The initial condition for $\Lambda_0 \to \infty$ is $\Sigma^{\Lambda_0} = 0$.\cite{KEM}
In the lowest-order scheme the two-particle vertex is the bare 
anti-symmetrized interaction and reads:
\begin{eqnarray*}
\Gamma_{a',b';a,b}&=&\left[U\left(1\hspace{-1pt}-\delta_{\sigma_a,\sigma_b} \right)\delta_{j_a,j_b}+U'\left(\delta_{j_a,j_b+1}\hspace{-1pt}+\delta_{j_a,j_b-1} \right) \right]\\
&& \hspace{-30pt}\times \left( \delta_{j_a',j_a}\delta_{\sigma_a',\sigma_a}\delta_{j_b',j_b}\delta_{\sigma_b',\sigma_b}-\delta_{j_a',j_b}\delta_{\sigma_a',\sigma_b}\delta_{j_b',j_a}\delta_{\sigma_b',\sigma_a} \right)\, .
\end{eqnarray*}
Dynamical contributions to $\Sigma^{\Lambda}$ are generated only at higher orders. As the latter are 
important for the conductance at finite temperatures $T>0$, the present approximation scheme 
is restricted to $T=0$. The correct temperature dependence of the (single-dot) Kondo 
ridge is only captured if the flow of a frequency dependent two-particle vertex---leading 
to a flowing frequency dependent self-energy---is kept.\cite{karrasch1,karrasch2,Severin}
Within our approximation, the matrix elements $\Sigma^{\Lambda=0}_{a',a}=\tilde{\Sigma}_{a',a}$ of the self-energy 
at the end of the flow can be interpreted as 
interaction-induced renormalizations to the noninteracting model parameters such 
as the SOI and the on-site energies.\cite{KEM} Furthermore entirely new matrix elements will be generated if permitted by symmetry.
The full propagator including interaction effects is determined via the Dyson equation 
(\ref{dyson}), from which various observables can be computed.\cite{KEM} 
While the renormalized effective single-particle energy levels as a function of the bare parameters will be discussed in detail in Sec.~\ref{sec:ESOI},
we here concentrate on the linear conductance. At $T=0$ current-vertex 
corrections vanish and the Kubo formula for the spin-resolved 
conductance assumes a generalized 
Landauer-B\"uttiker form\cite{Oguri} 
\begin{equation*}
G_{\sigma,\sigma'} =\frac{e^2}{h} 
 \left|\mathcal{T}_{\sigma,\sigma'}(0)\right|^2 \; ,
\end{equation*}
with the effective transmission $\mathcal{T}_{\sigma,\sigma'}(0)$ evaluated 
at the chemical potential.  For the present setup the transmission is given by the matrix elements of the full propagator leading to\cite{KEM} 
\begin{eqnarray*}
G=\sum_{\sigma,\sigma'} G_{\sigma,\sigma'}& =& \frac{e^2}{h}4 \sum_{\sigma,\sigma'} \left| \sum_{j,j'}t_{Lj} t_{Rj'}\,\mathcal{G}_{j,\sigma;j',\sigma'}(0)\right|^2 \hspace{-6pt}.
\end{eqnarray*}

\section{Results}\label{sec:res}

\begin{figure}[t!]
\center{\includegraphics[clip=true,width=7cm]{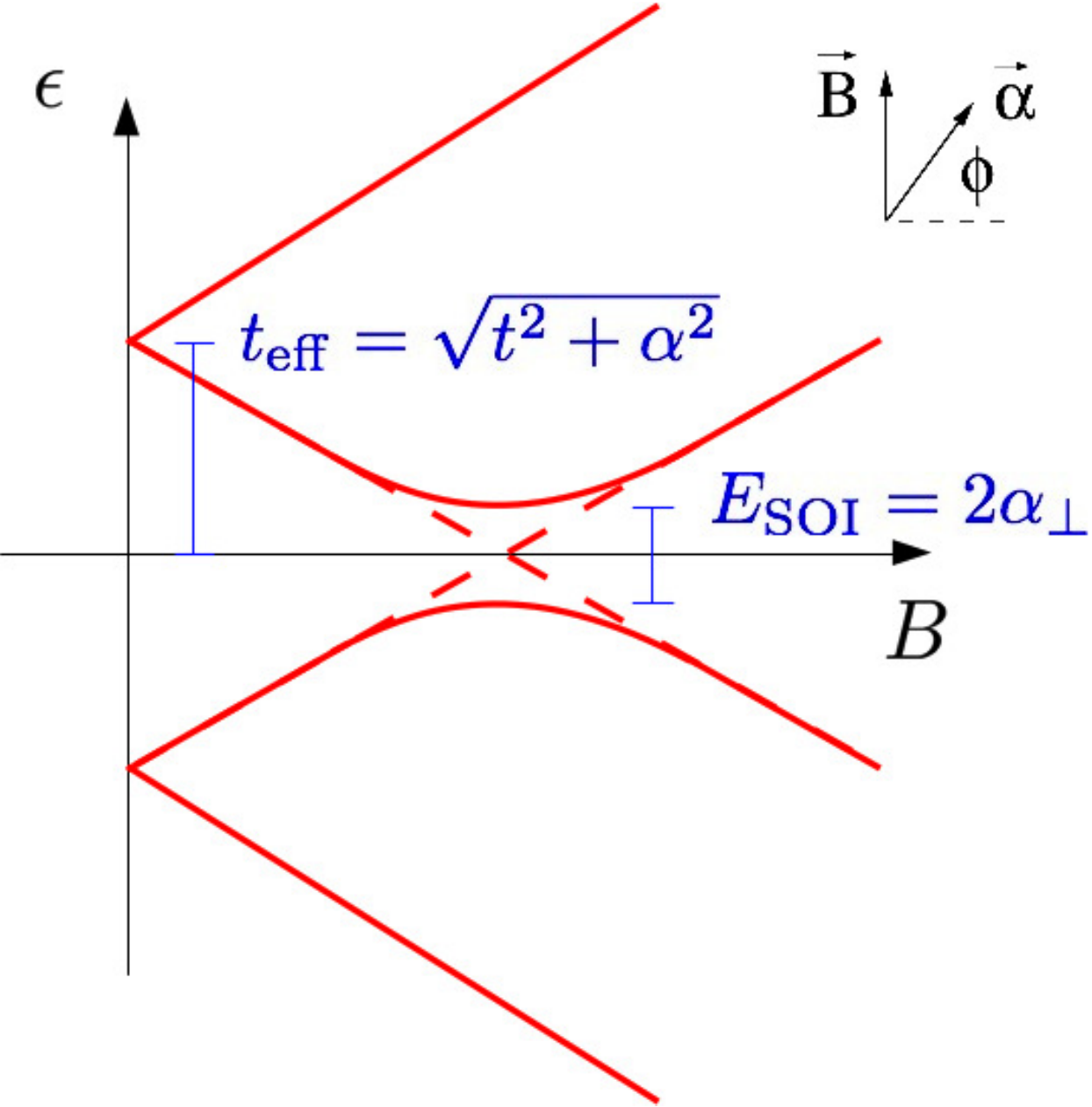}}
\caption{\label{fig:fig2} (Color online)
 Schematic representation of the noninteracting level structure of the isolated dot as relevant in Sec.~\ref{sec:ESOI}.}
\end{figure}

\subsection{Linear conductance}\label{sec:Cond}
The linear (and finite bias) transport characteristics allow to access the physics of quantum dots, and are of of particular interest in view of the use of dot setups as information processing devices.\cite{kanai,deacon,defranceschi,preprint} In experiments large ranges of applied gate voltages and Zeeman fields as well as different orientations of the field can be analyzed. The fRG was shown to capture the effects of the two-particle interaction on the linear conductance 
for general models\cite{grap,KEM} as well as 
for models specifically tailored to describe experimental setups.\cite{preprint} 
We here compute the conductance for various parameters of our  minimal model to study multi-level dots with sizeable SOI as realized in experiments. As using experimental parameters without a well established microscopic model is difficult, we will choose our parameters in the following in such a way as to facilitate the discussion. As a guide for the magnitude of the single particle parameters we use Ref.~[\onlinecite{preprint}].
\begin{figure}[t!]
\center{\includegraphics[clip=true,width=8cm]{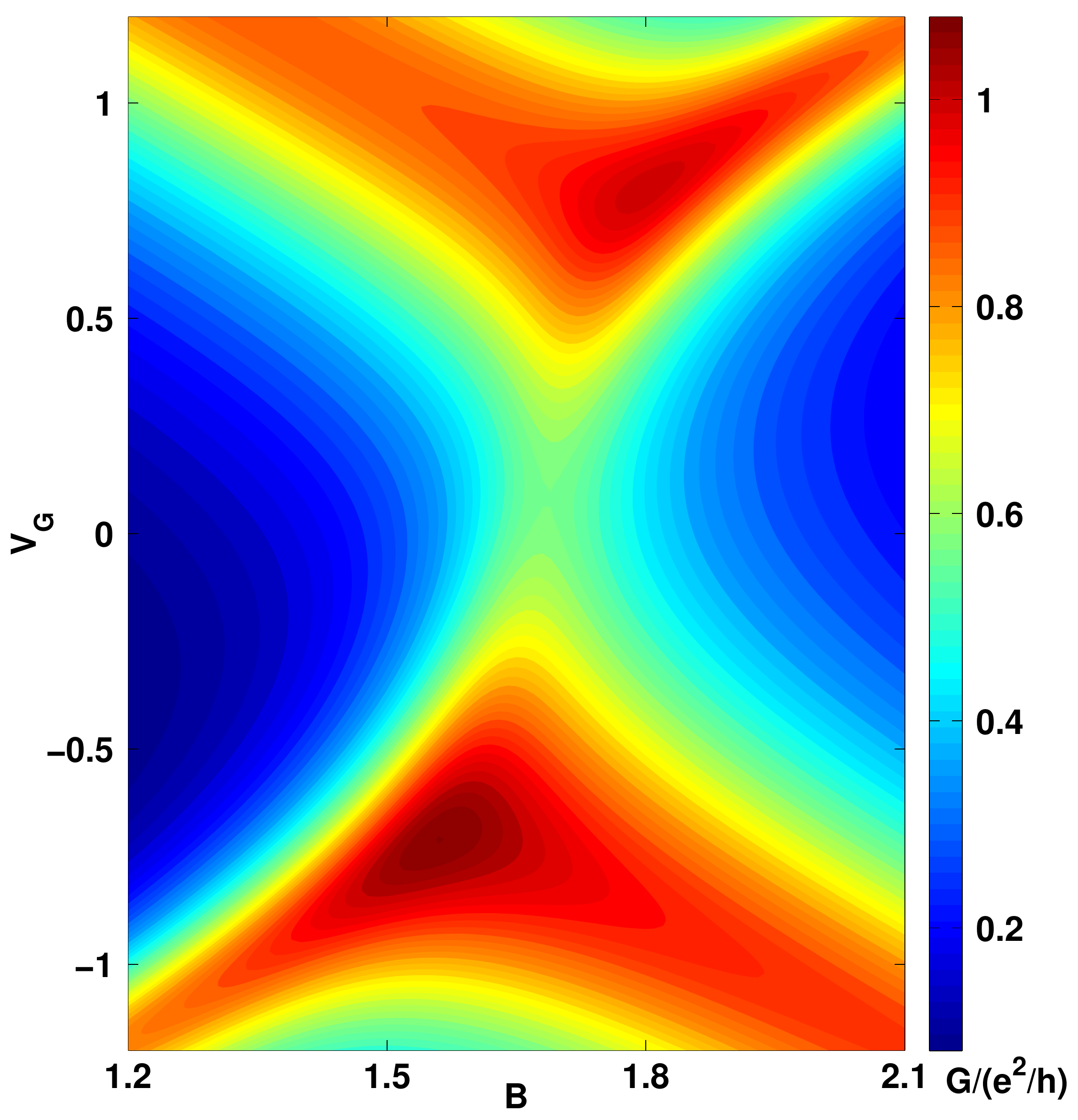}}
\caption{\label{fig:fig3} (Color online)
Qualitative reproduction of the conduction region of  Fig.~1c in Ref.~[\onlinecite{kanai}]. The chosen system parameters are: $t=\alpha=1$, $\delta=0.9$, $\phi=0.46\pi$, $U=U^{\prime}=1.5$, $t_{L,1}=-0.5$, $t_{L,2}=0.45$, $t_{R,1}=0.4$, $t_{R,2}=-0.3$.\cite{FootnoteCouplingsII}}
\end{figure}
The noninteracting level structure of the isolated dot is schematically shown in Fig.~\ref{fig:fig2} for $\epsilon_1=\epsilon_2=V_{\rm G}=0$ and can be utilized to obtain a rough picture of the linear conductance. 
The approximate fRG conserves particle-hole symmetry 
translating into a symmetric linear conductance with respect to the gate voltage transformation $V_{\rm G}\rightarrow -V_{\rm G}$.\cite{KEM}
For vanishing Zeeman field $B=0$ there are two well separated spin degenerate energy levels at $\epsilon=\pm t_{\rm eff} $. When those levels cross the chemical potential of the leads, the Kondo effect 
will lead to the characteristic conductance plateaus of height $G_{\rm Max}=2e^2/h$ (for a totally symmetric serial dot).\cite{grap} The width of the plateaus is determined by the local Coulomb interaction. In presence of a finite field $B$ parallel to the spin-orbit direction additional spin degenerate levels occur at $V_{\rm G}=0,\ B_z=\pm t_{\rm eff}$ (dashed lines in Fig.~\ref{fig:fig2}) giving rise to Kondo correlations. Similarly to the symmetry in the gate voltage, the conductance is also invariant under $B\rightarrow -B$. This symmetry also holds in the following more general cases.  If the SU(2) spin symmetry is broken by SOI (defining the $z$-direction in the spin space, see Eqs.~(\ref{HSOI}) and (\ref{HZ})) this finite-$B$ degeneracy can be lifted via a Zeeman field component perpendicular to the SOI and the resulting anti-crossing (full lines in Fig.~\ref{fig:fig2}) suppresses the Kondo effect on an exponential scale. 
For different coupling strengths to the leads the maximum conductance on the Kondo plateau is reduced by a factor which depends on the dot parameters. Including additional asymmetries in the on-site energies $\epsilon_1- \epsilon_2=2\, \delta\neq0$ gives rise to finite-field Kondo ridges bent with respect to the $V_{\rm G}$ axis.\cite{grap,preprint} Aside from the development of the Kondo effect due to the local Coulomb interaction,  the nearest-neighbor interaction renormalizes the position of conductance resonances.
 
With this simplified multi-level quantum dot model we can provide a good qualitative  
description of the various parameter regimes of recent experiments. To exemplify this in Fig.~\ref{fig:fig3} we show an asymmetrically coupled dot with different on-site energies in a Zeeman field with a perpendicular component to the spin-orbit direction. The results feature a reduced conductance and a slanted resonance with a pronounced suppression due to the anti-crossing of the levels, both characteristic for the asymmetric setup. For the chosen parameters (see the caption) the $B$- and $V_{\rm G}$- dependence of $G$ strongly resembles the experimental data shown in Fig. 1c of Ref.~[\onlinecite{kanai}].

In the following we use the minimal multi-level model to compute the gap in the single-particle spectrum (see Fig.~\ref{fig:fig2}) for different relative orientations of the SOI and the Zeeman field as recently investigated experimentally.\cite{kanai,deacon,defranceschi} A similar level splitting for small Zeeman fields close to the $B\simeq 0$ Kondo ridges allows to determine the effective $g$-factor $g_{\rm Levels}$.
Alternatively, $g_{\rm Cond}$ can be extracted from the Coulomb blockade peak splitting of the conductance.
For both we will compare our theoretical results to the experimental ones.
In order to disentangle the effects under consideration from asymmetry effects (possibly enhanced or suppressed by the interaction) we will focus on a serial double quantum dot geometry. Detailed results for the conductance of this model were presented in Ref.~\onlinecite{grap}.

\subsection{Spin orbit energy $E_{\rm SOI}$}\label{sec:ESOI}
\begin{figure}[t!]
\center{\includegraphics[clip=true,width=8.5cm]{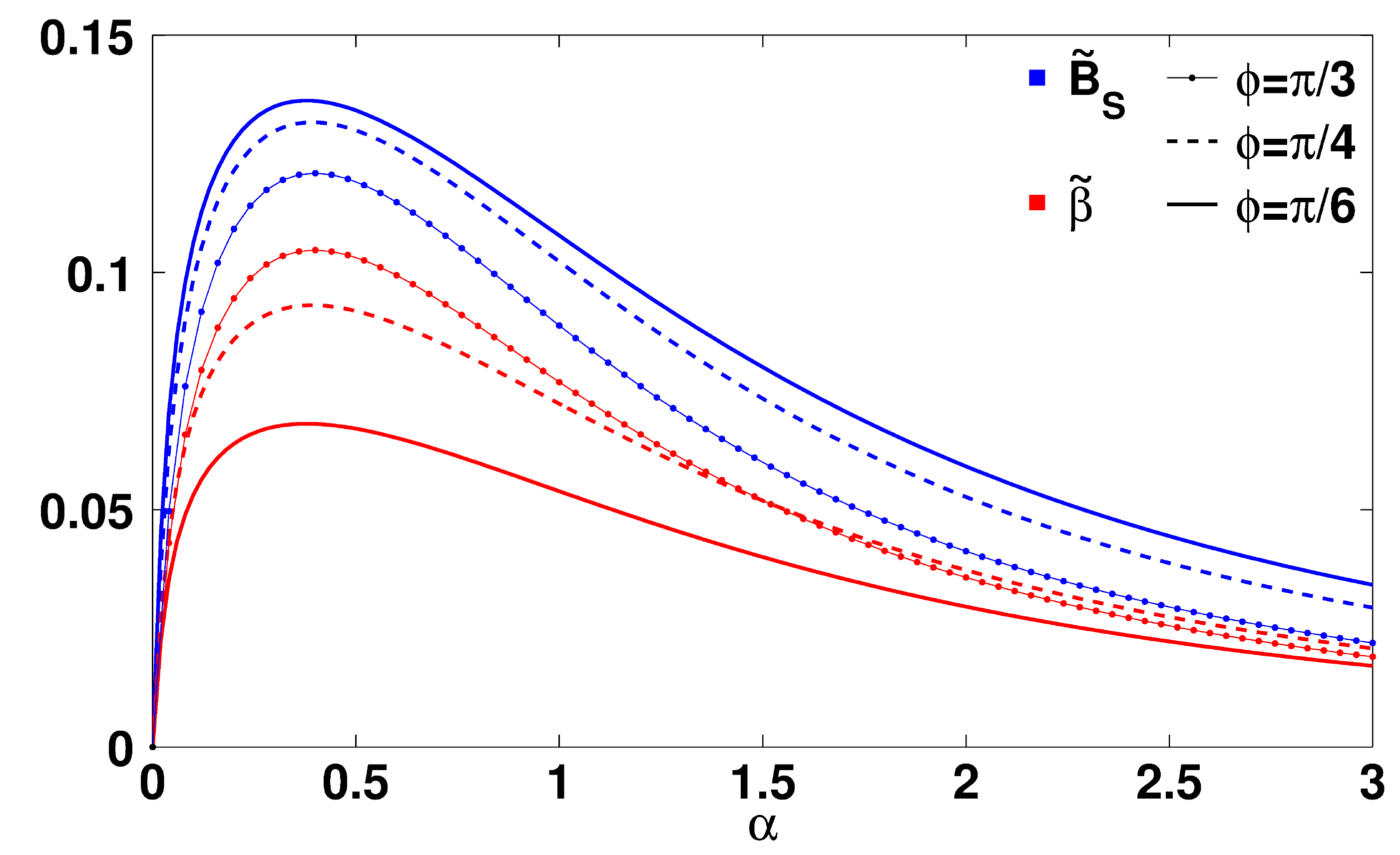}}
\caption{\label{fig:fig4} (Color online)
Bare spin-orbit parameter $\alpha$ dependence of the generated effective single-particle parameters $\tilde{B}_S$ (upper (blue) curves) and $\tilde{\beta}$ (lower (red) curves) for different orientations of the bare magnetic field (full lines $\phi=\frac{\pi}{6}$, dashed lines $\phi=\frac{\pi}{4}$, dotted lines $\phi=\frac{\pi}{3}$). The remaining parameters are: $t=1$, $\delta=0$, $B=1.2$, $U=U^{\prime}=1$, $t_{\rm Coup}=0.4$.}
\end{figure}
Recent experiments on InAs devices, where SOI is relevant, showed Kondo-like features in the linear conductance at finite Zeeman field.\cite{kanai,takahashi} Cotunneling spectroscopy allows to resolve the dependence of the involved energy levels on the $B$ field and its orientation relative to the SOI direction.~\cite{Fussnote}
This analysis shows that any finite orthogonal component of the Zeeman fields lifts the degeneracy of the two states responsible for the Kondo plateau. For the generic situation an anti-crossing of these two states is observed, varying periodically with the orientation of the Zeeman field. The minimal size of the gap as a function of the magnitude of the Zeeman field defines the energy $E_{\rm SOI}$ (see Fig.~\ref{fig:fig2}). 
For the theoretical description we consider the symmetrically coupled serial double-dot geometry at vanishing gate voltage and level splitting - i.e. $\epsilon_1=\epsilon_2=0$, $t_{L,1}=t_{R,2}=t_{\rm Coup}$ and $t_{L,2}=t_{R,1}=0$ in Fig.~\ref{fig:fig1}. This implies $\Gamma_1=\Gamma_2=\Gamma$. At any (fixed) cutoff value during the RG flow the single-particle part of our system (omitting the lead terms of Eqs.~(\ref{lead}) and (\ref{Hcoup})) can be described by the following Hamiltonian in the basis of Eq.~(\ref{basis}):
\begin{eqnarray*}
\hspace{-0.1cm}h&=&\begin{pmatrix}
 B_z	&B_x+iB_S& -t+i\alpha&-i\beta	\\
B_x-iB_S& -B_z& -i\beta	&-t-i\alpha	\\
-t-i\alpha	&i\beta&B_z&B_x-iB_S	\\
i\beta& -t+i\alpha&B_x+iB_S&-B_z	\\
 \end{pmatrix}\;,
\end{eqnarray*}
with all matrix elements depending on the cutoff $\Lambda$.
\begin{figure}[t!]
\center{\includegraphics[clip=true,width=8.5cm]{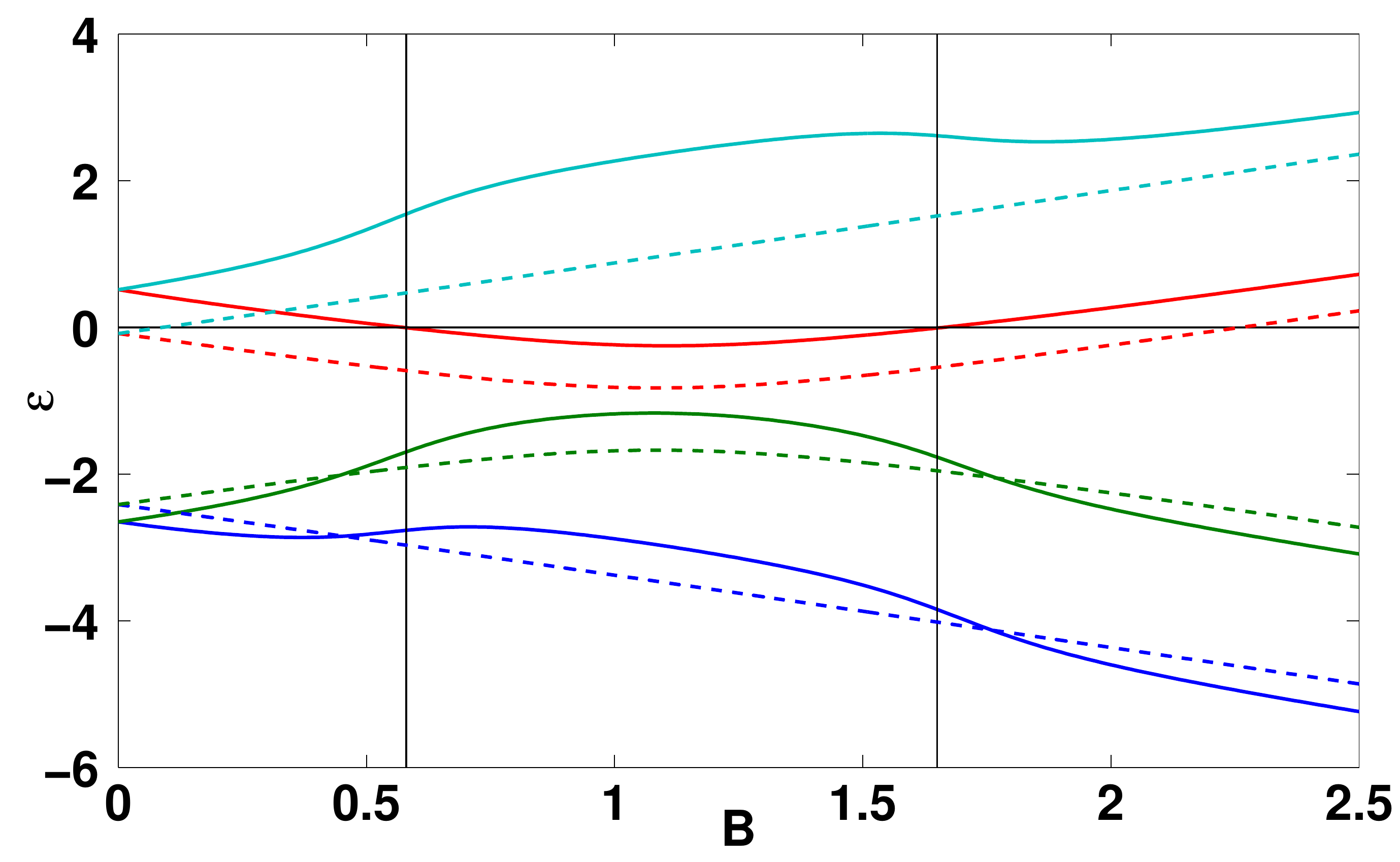}}
\caption{\label{fig:fig5} (Color online)
 Bare Zeeman field dependence of the effective single-particle energy spectrum (full lines) as well as of the corresponding noninteracting ones (dashed lines). As a guide to the eye the chemical potential and the crossing position of the renormalized levels are given as straight black lines. The parameters are: $V_{\rm G}=1.25$, $t=1$, $\alpha=0.6$, $\phi=0.25\pi$, $U=U^{\prime}=1$, $t_{\rm Coup}=0.4$.}
\end{figure}
The initial conditions are given by the bare values of the noninteracting system, with $B_{z}=B\sin{\phi}$ and $B_{x}=B\cos{\phi}$. The Hamiltonian (matrix) $h$ contains the parameters $B_S$ and $\beta$ which are zero initially but are generated by the two-particle interaction during the RG flow. The new parameter $B_S$ is a Zeeman field perpendicular to both the applied Zeeman field and the SOI direction, with opposite orientation on the two dot sites, while the spin flip hopping $\beta$ is a Dresselhaus SOI term.\cite{Winkler}
The appearance of an effective Zeeman field induced by a finite Coulomb interaction in presence of a broken spin symmetry is discussed for quantum dots with ferromagnetic leads\cite{zitko0,koenig} and has been observed recently in systems involving SOI.\cite{jens,nowak}
 We find a non-monotonic dependence of the effective $\tilde{B}_S=B_S^{\Lambda=0},\, \tilde{\beta}=\beta^{\Lambda=0}$ for varying the initial value of the SOI - given by $\alpha$ - and initial relative orientation of the Zeeman field and the SOI parametrized by the angle $\phi$ as shown in Fig.~\ref{fig:fig4}.

At the end of the flow the effective parameters include the renormalization of the initial values due to the Coulomb interaction. Here we focus on the angular dependence.
For the considered setup the flow equations assume the convenient analytical form in terms of vectors $\vec{B}_{\rm e}=(B_x,B_z,B_S)^{T}$ and $\vec{t}_{\rm e}=(\alpha,\beta,t)^{T}$
\begin{eqnarray}
\label{ESOIFlow}
\dot{\vec{B}}_{\rm e}&=&-\frac{U}{\pi D(\Lambda)}\left[2 \left( \vec{t}_{\rm e} \cdot \vec{B}_{\rm e}\right)\vec{t}_{\rm e}+f_+ \vec{B}_{\rm e} \right]\nonumber\\
\dot{\vec{t}}_{\rm e}&=&-\frac{U^{\prime}}{\pi D(\Lambda)}\left[2 \left( \vec{t}_{\rm e} \cdot \vec{B}_{\rm e}\right)\vec{B}_{\rm e}+f_- \vec{t}_{\rm e} \right] \;,
\end{eqnarray} 
where we introduced
\begin{eqnarray*}
f&=&|\Lambda|+\Gamma \ ,\  f_{\pm}=f^2\pm\left( |\vec{B}_{\rm e}|^2-|\vec{t}_{\rm e}|^2\right)\;,\\
D(\Lambda)&=&\det(if-h)=f_+^2+2f^2|\vec{t}_{\rm e}|^2+4\left( \vec{t}_{\rm e} \cdot \vec{B}_{\rm e}\right)^2\ .
\end{eqnarray*}
As a consequence, $B_S$ is generated only for both finite initial $B_{x}$ and $\alpha$
(compare to Fig.~\ref{fig:fig4}). 
For the hopping $\beta$ to be generated, a non-vanishing $B_{z}$ is additionally required. 
\begin{figure}[t!]
\center{\includegraphics[clip=true,width=8.5cm]{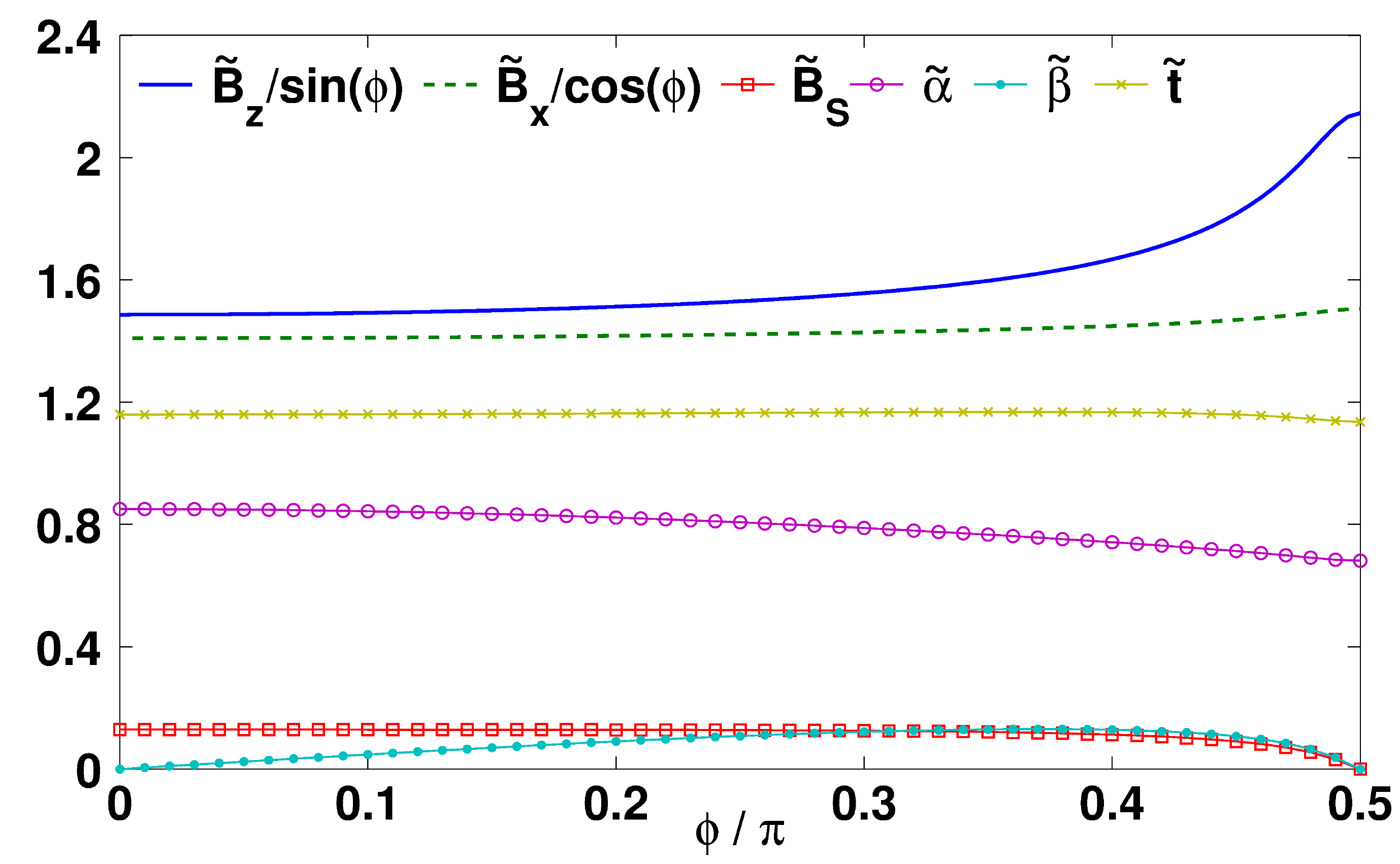}}
\caption{\label{fig:fig6} (Color online)
 Renormalization-induced $\phi$ dependence of the effective parameters. Here we compensated for the bare angular dependence of the Zeeman fields by dividing by the corresponding trigonometric function. The initial parameters are: $t=1$, $\alpha=0.6$, $B=1.2$, $U=U^{\prime}=1$, $t_{\rm Coup}=0.4$.}
\end{figure}

\begin{figure}[t!]
\center{\includegraphics[clip=true,width=8.5cm]{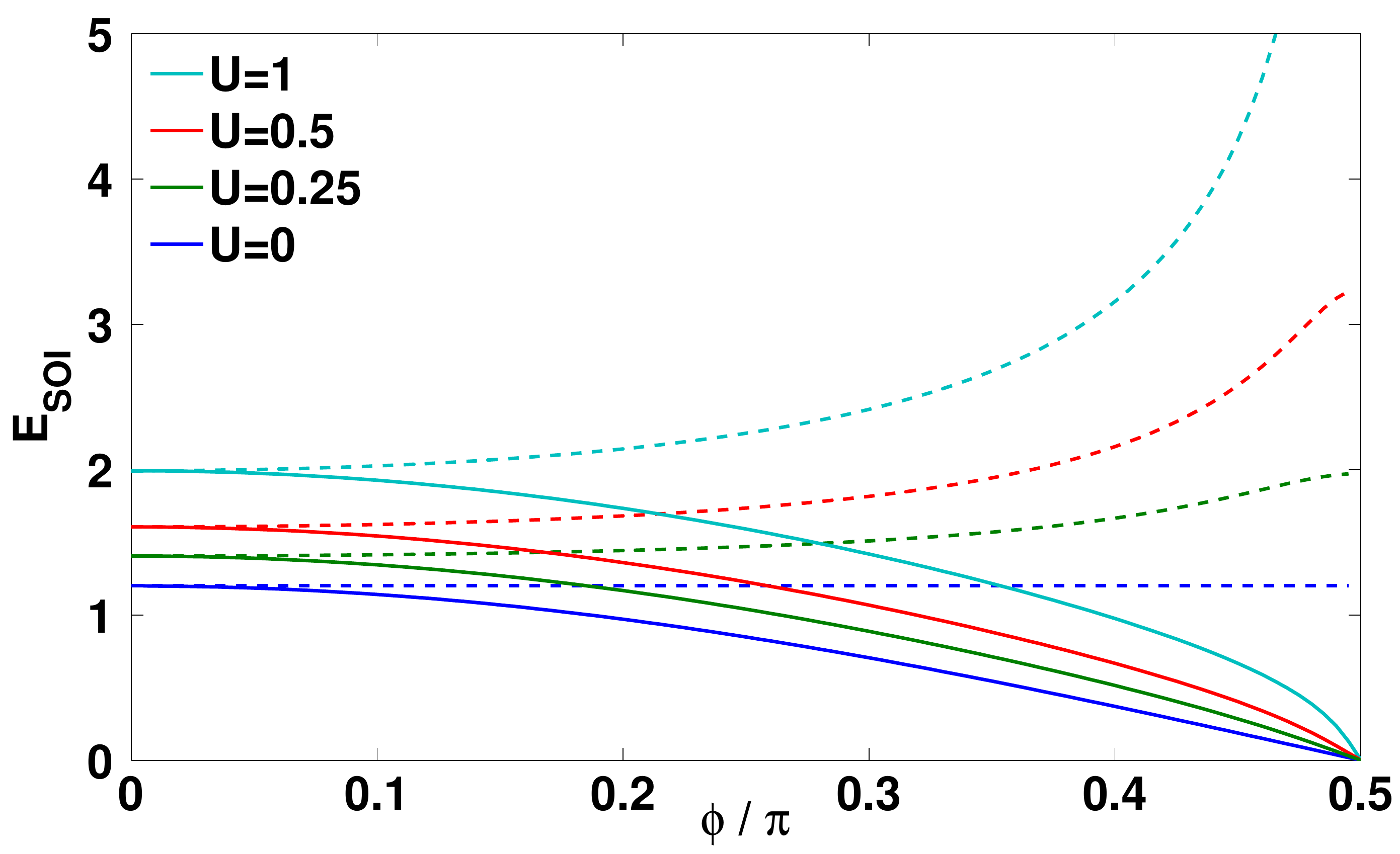}}
\caption{\label{fig:fig7} (Color online)
Spin-orbit energy versus orientation of the Zeeman field for several values of the interaction $U=U^{\prime}$(full lines). The dashed lines show the spin-orbit energies normalized to the angular dependence of the noninteracting case. The initial conditions are given by $t=1,\ \alpha=0.6,\ t_L=t_R=0.3$.}
\end{figure} 
 Diagonalizing the Hamiltonian $h$ yields the eigenvalues (the symmetric leads contribute a constant imaginary part which is omitted here, compare Sec.~\ref{sec:Asym}):
\begin{equation*}
\epsilon=\pm\sqrt{|\vec{t}_{\rm e}|^2+|\vec{B}_{\rm e}|^2\pm2\sqrt{|\vec{B}_{\rm e}|^2|\vec{t}_{\rm e}|^2-\left( \vec{t}_{\rm e} \cdot \vec{B}_{\rm e}\right)^2}}\; .
\end{equation*}
They can be interpreted as the single-particle levels of a corresponding noninteracting system. This interpretation has already been applied successfully to the problem of phase lapses in multi-level quantum dots.\cite{PhaseLapsA,PhaseLapsB} 
Figure \ref{fig:fig5} shows data for finite gate voltages for which pronounced renormalization effects occur if one of the involved levels crosses the leads' chemical potential. For this situation the other energy levels are shifted upwards due to charging effects. This is not observed at $V_{\rm G}=0$ as no level crosses the chemical potential, but the overall shift of the effective levels with respect to the noninteracting ones apparent in Fig.~\ref{fig:fig5} remains and will be of importance in the following. From the two intermediate levels we determine the spin-orbit energy 
\begin{multline*}
E_{\rm SOI}=\\
{\rm min}_{B}\left(\hspace{-2pt}2\sqrt{|\vec{t}_{\rm e}|^2\hspace{-2pt}+\hspace{-2pt}|\vec{B}_{\rm e}|^2\hspace{-2pt}-2\sqrt{|\vec{B}_{\rm e}|^2|\vec{t}_{\rm e}|^2\hspace{-2pt}-\hspace{-2pt}\left( \vec{t}_{\rm e} \cdot \vec{B}_{\rm e}\right)^2}}\right)\,.
\end{multline*}
The bare initial value $E_{\rm SOI}=2\,\alpha \,\cos{\phi}=2\,\alpha_{\perp}$ is obtained for $B^2_{\rm min}=t^2+\alpha_{||}^2$, with $\alpha_{||}=\alpha \,\sin{\phi}$.
Due to the complicated non-linear structure of the flow equations an analytic expression of the renormalized spin-orbit energy in terms of the bare Zeeman field can not be obtained.  
The numerical solution of Eq.~(\ref{DGLsigma}) or (\ref{ESOIFlow}) shows that the renormalized effective parameters acquire a non-trivial angular dependence, as seen in Fig.~\ref{fig:fig6}.
 The results depend only quantitatively on the details of the Coulomb interaction 
and the inter-dot hopping $t$, as long as  $U^{(\prime)}/\Gamma$ and $t/\Gamma$ are 
sufficiently large. We here focus on $U=U'$ and $t=1$. 
The renormalized spin-orbit energies are shown as solid lines in Fig.~\ref{fig:fig7} and follow the general form of the bare case calculated above: We find a maximum of the level splitting for the perpendicular orientation $\phi=0$ and a monotonous decrease to $E_{\rm SOI}=0$ when increasing $\phi$ towards the parallel configuration at $\phi=\pi/2$. In a detailed examination for more values of the interaction the maximum is seen to increase linearly with $U$. The similarity of the interacting and noninteracting curves might lead to the expectation that the maximum is given by twice the renormalized SOI parameter $2\,\tilde{\alpha}$, but this is not the case. The dashed lines in Fig.~\ref{fig:fig7} show the renormalized spin-orbit energies divided by the angular dependence of the bare case, i.e. $\cos{\phi}$. The strong deviations from this bare dependence close to the parallel orientation leads to the important result that the functional dependence of the SOI energy on $\phi$ is strongly affected by the two-particle interaction.
This can be traced back to the more pronounced renormalization effects of the bare parameters around the parallel configuration as is seen in Fig.~\ref{fig:fig6}, while the parameters are mostly unaffected for $\phi\lessapprox 0.3\pi$. This behavior can be intuitively understood in the following way: for the parallel configuration a sufficiently strong Coulomb interaction gives rise to the finite-$B$ Kondo effect related to the level crossing. With the vanishing Kondo effect in presence of a finite orthogonal $B$-field component relative to the SOI the Coulomb interaction effects appear suppressed as well.
We finally note that our results agree qualitatively with the experimental results of Refs.~[\onlinecite{kanai}] and [\onlinecite{takahashi}].

\subsection{Effective $g$-factors}\label{sec:g}
\begin{figure}[t!]
 \center{\includegraphics[clip=true,width=8.5cm]{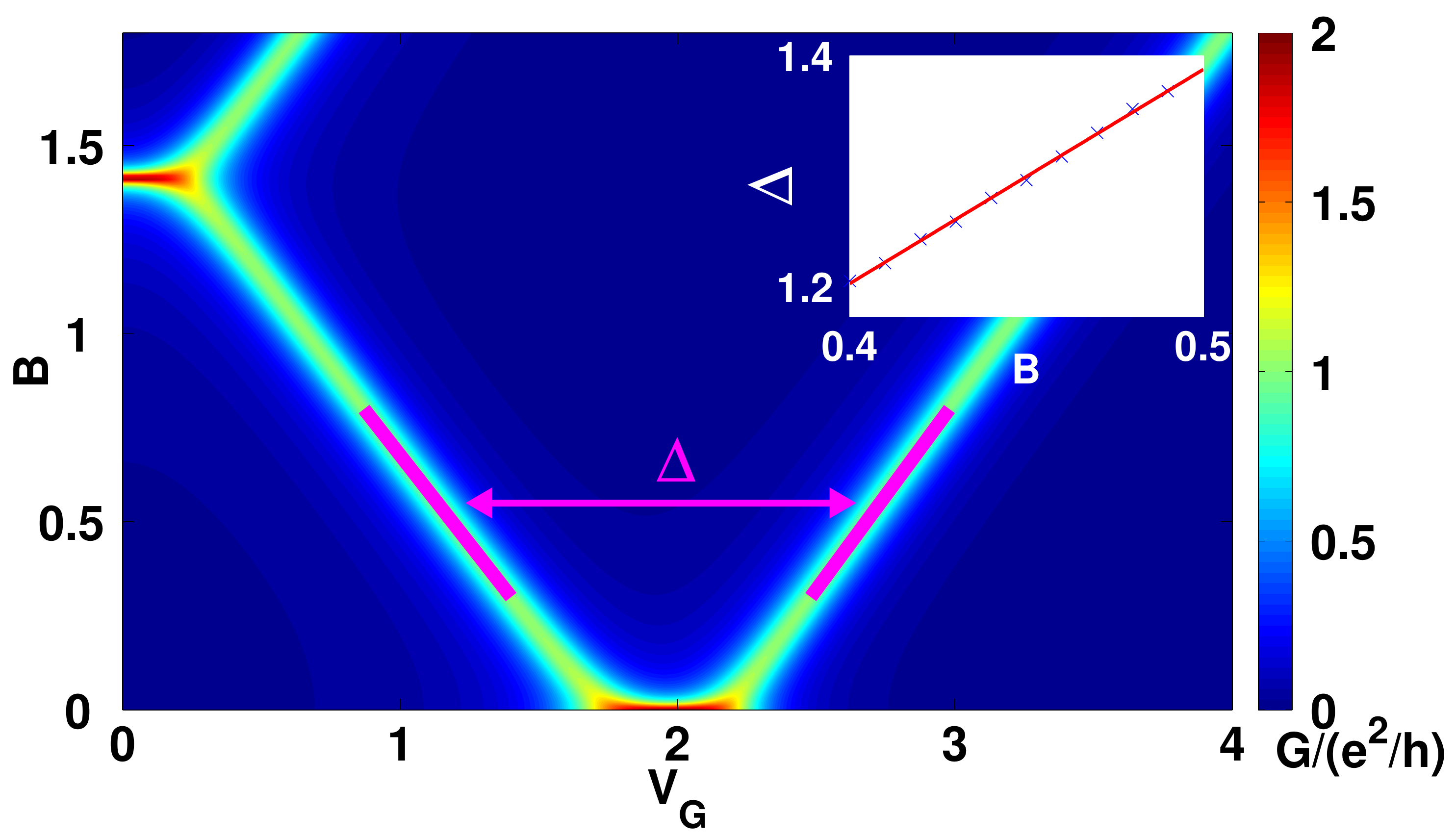}}
 \caption{\label{fig:fig8} (Color online)
  Determination of the $g$-factor $g_{\rm Cond}$ from the gate-voltage dependence 
  of the Coulomb blockade peaks in the linear conductance. Bare parameters: $t=1$, $\alpha=1$, $\phi=0.5\pi$, $U=U^{\prime}=0.5$, $t_{\rm Coup}=0.3$. The inset shows a linear fit to the extracted gate voltage difference $\Delta$.}
 \end{figure}
Experimentally the $\phi$-dependence of effective $g$-factors is studied as well. In this section we will model two different protocols used for their extraction. In the first one, following the experiments of 
Refs.~[\onlinecite{kanai}], [\onlinecite{deacon}], and [\onlinecite{nilsson}] the Coulomb blockade peak splitting around a $B=0$ Kondo resonance is extracted from the linear conductance defining $g_{\rm Cond}$. As the fRG reliably reproduces the linear conductance this procedure is easily adopted. In the second protocol bias spectroscopy is used to measure the level splitting in the vicinity of a Kondo resonance, determining $g_{\rm Levels}$. 
\subsubsection*{{\bfseries{Effective $g$-factors: $g_{\rm Cond}$}}}
\begin{figure}[t!]
 \center{\includegraphics[clip=true,width=8.5cm]{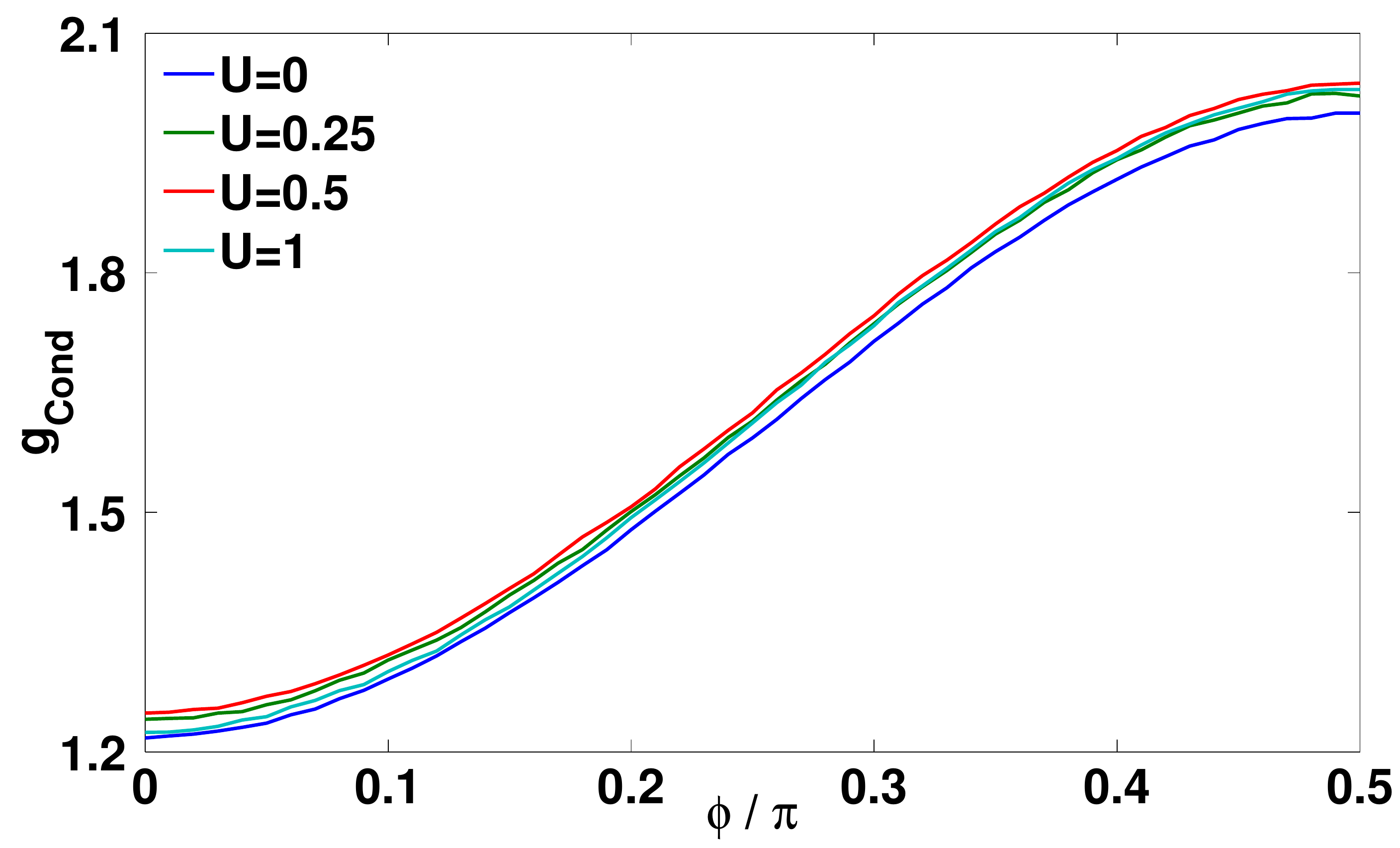}}
 \caption{\label{fig:fig9} (Color online)
  Zeeman field orientation dependence of $g_{\rm Cond}$ as extracted from the gate-voltage dependence 
  (see Fig.~\ref{fig:fig8}) for different values  of the Coulomb interaction $U=U'=0,0.25,0.5,1$ and
  parameters as in Fig.~\ref{fig:fig8}.}
 \end{figure}

As seen in linear conductance measurements for $B>0$ resonance peaks of maximal height $e^2/h$ and 
width $\sim\Gamma$ develop out of Kondo plateaus, corresponding to the filling of a dot state. From a linear fit to the splitting of these Coulomb blockade peaks at small to intermediate Zeeman fields we determine $g_{\rm Cond}$ by identifying it with the slope of the fit as shown in Fig.~\ref{fig:fig8}, see i.e. the experiments reported in Ref.~[\onlinecite{deacon}], Fig. 1c. 
Due to the presence of the $B=0$ Kondo ridge, 
there is an offset before the linear behavior sets in. 
The full angular dependence of $g_{\rm Cond}$ in the interacting system is shown in Fig.~\ref{fig:fig9},
and for comparison also the noninteracting result is displayed. No significant effects of the two-particle interaction are observed. For all values of the interaction, $g_{\rm Cond}$ exhibits an S-shaped dependence on the relative orientation of the Zeeman field and the SOI with a maximum for the parallel configuration and a finite minimum at $\phi=0$.
\subsubsection*{{\bfseries{Effective $g$-factors: $g_{\rm Levels}$}}}
In order to use the fRG to model the effective $g$-factor $g_{\rm Levels}$ we will make use of the effective-level interpretation, according to the discussion in the previous section.~\cite{Fussnote} Introducing a finite gate voltage as additional parameter, an explicit analytic form of the flow equation as for the spin-orbit energy (see Eq.~(\ref{ESOIFlow})) is much more difficult to obtain. Thus we will solve the general flow equation Eq.~(\ref{DGLsigma}) including a numerical inversion of the matrix on the right hand side of Eq.~(\ref{dyson}) and extract the low field splitting from the eigenvalues of the resulting effective Hamiltonian. 
As for $g_{\rm Cond}$, we follow the experimental procedure~\cite{katsaros}, perform a linear fit to the computed splitting and identify the slope as $g_{\rm Levels}$. In principle we can extract a gate-voltage dependence for this quantity but here we choose the gate voltage such that the linear conductance around the $B=0$ Kondo resonances is maximal.
We note that in the single-impurity Anderson model $g_{\rm Levels}$ can be related to the magnetic susceptibility.\cite{yamada}

\begin{figure}[t!]
\center{\includegraphics[clip=true,width=8.5cm]{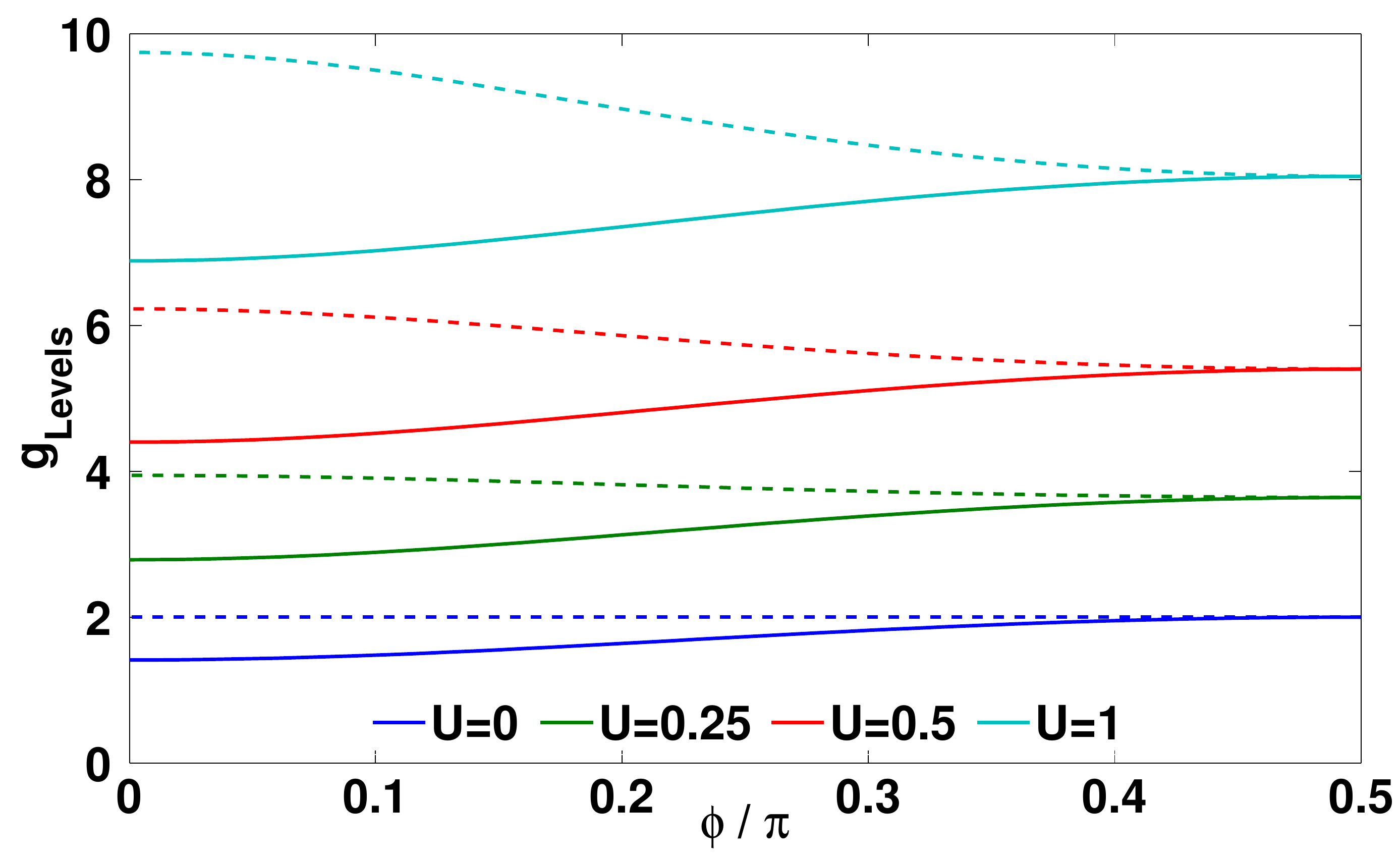}}
\caption{\label{fig:fig10} (Color online)
 Angular dependence of $g_{\rm Levels}$ as extracted from the level splitting (solid lines)
 for the same parameters as in 
 Fig.~\ref{fig:fig8} and different values
 of the Coulomb interaction $U=U'=0,0.25,0.5,1$. In addition, the angular dependence normalized to the noninteracting case is shown (dashed lines).}
\end{figure}

The noninteracting Hamiltonian with $\epsilon_1=\epsilon_2=V_{\rm G}$ yields the eigenvalues
\begin{eqnarray}
\epsilon& =&V_{\rm G}\pm\sqrt{B^2+t^2+\alpha^2\pm2B\sqrt{t^2+\alpha_{||}^2}}\nonumber\\
&=&V_{\rm G}\pm\sqrt{\left(B\pm\sqrt{t^2+\alpha_{||}^2}\right)^2+\alpha_{\perp}^2}\nonumber
\label{e}
\end{eqnarray}
from which the bare $g_{\rm Levels}$ is identified
\begin{equation*}
 |\Delta \epsilon^{\pm}|\approx 2\frac{\sqrt{t^2+\alpha^2_{||}}}{t_{\rm eff}}B=g_{\rm Levels} B\;, \;\;\mbox{ for } B\ll 1\;.
\end{equation*}
The angular dependence for the symmetric case at fixed $V_{\rm G}$ 
is shown in Fig.~\ref{fig:fig10} for different values of the Coulomb interaction. The general form is again S-shaped as for $g_{\rm Cond}$. In contrast to $g_{\rm Cond}$, for $g_{\rm Levels}$ we find similar interaction effects as for $E_{\rm SOI}$. Again the renormalization of the maximum value of $g_{\rm Levels}$ is a prominent effect. Comparing the $\phi$-dependence of the interacting curves with the noninteracting ones we find deviations which are much less pronounced compared to the ones of the spin-orbit energy (compare dashed lines of Figs.~\ref{fig:fig10} and \ref{fig:fig7}). Even though the qualitative $\phi$-dependence is similar to the one of $g_{\rm Cond}$, the strong renormalization of the amplitude implies that $g_{\rm Levels}\neq g_{\rm Cond}$. This makes it necessary to clearly distinguish between these two quantities as well as other contextually similar definitions using different computation methods or extraction protocols.~\cite{moore,costi,konik2,quay,zitko}

\subsection{Asymmetry effects}\label{sec:Asym}
\begin{figure}[t!]
\center{\includegraphics[clip=true,width=8.5cm]{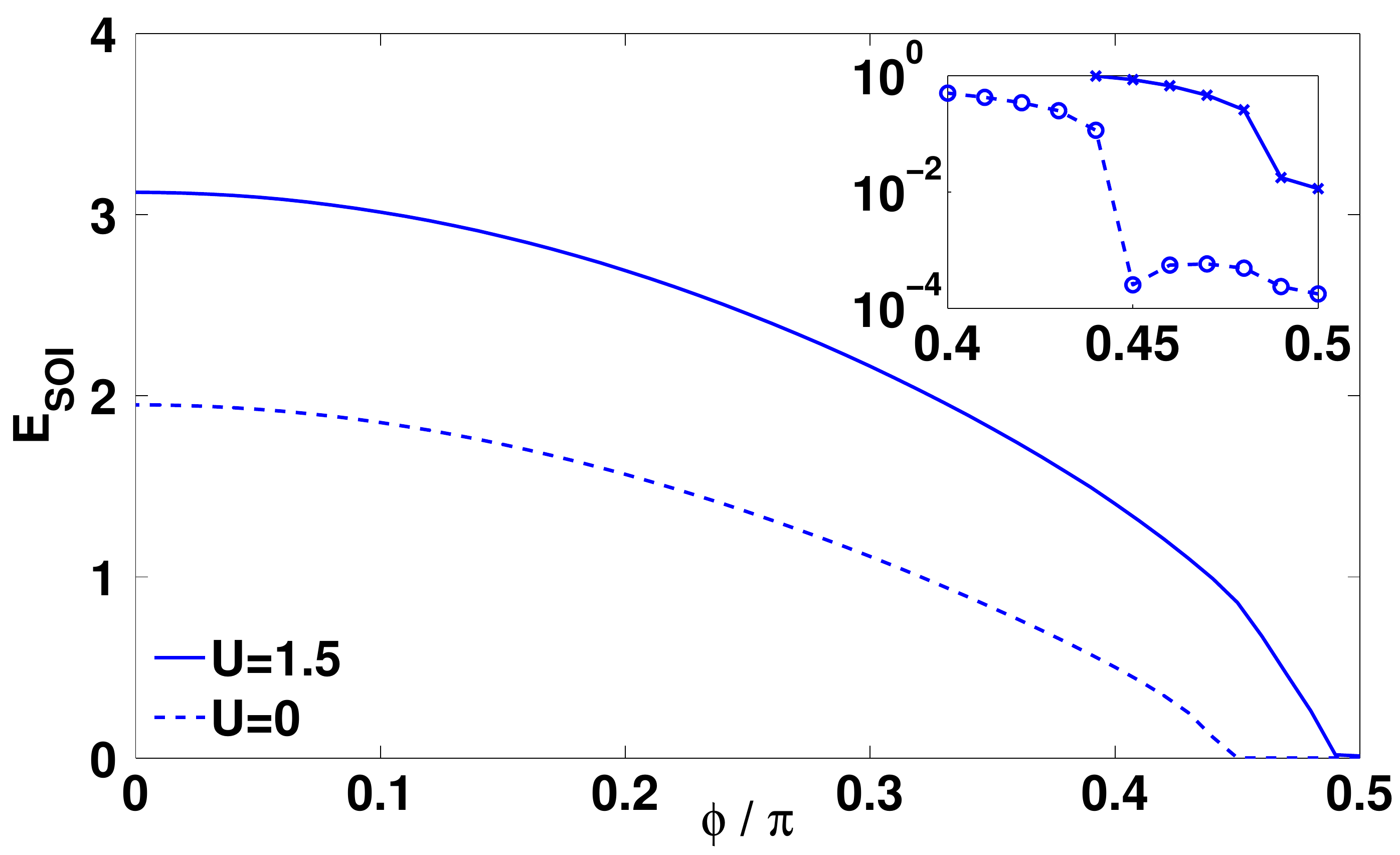}}
\caption{\label{fig:fig11} (Color online)
 Angular dependence of the spin-orbit energy for the parameters of Fig.~\ref{fig:fig3} (full lines). For reference, the angular dependence of the noninteracting case is shown (dashed lines). The inset shows a zoom in of the $\phi >0.4\pi$ range, note the logarithmic scale on the $y$-axis.}
\end{figure}
Experimental setups are more accurately modeled if we consider the more general situation of four (asymmetric) couplings and different on-site potentials as depicted in Fig. 1. Conceptually and computationally this is easily implemented within the fRG approach but due to the multitude of parameters an in depth analysis is beyond the scope of this work. Instead we will use the experimentally motivated parameters of Fig. 3 and calculate the spin-orbit energy and effective $g$-factors for this asymmetric setup to illustrate asymmetry effects. 
\subsubsection*{{\bfseries{Spin orbit energy $E_{\rm SOI}$}}}
While the calculation of the $g$-factors remains unaffected by the setup, the protocol for the spin-orbit energy has to be slightly expanded. In the experiments the parameters were tuned in such a way that the dot was at half filling to insure maximum degeneracy. In Sec.~\ref{sec:ESOI} this half filling condition was guaranteed for $V_{\rm G}=0$ due to particle-hole symmetry. In the considered asymmetric setup and for finite two-particle interaction this does not hold necessarily and we will fix the gate voltage and Zeeman field amplitude so that on average two electrons occupy the dot and the considered level splitting $E_{\rm SOI}$ is minimal. For the noninteracting case the dot is still half filled at $V_{\rm G}=0$ and the spin-orbit energy can be computed after diagonalizing Eq.~(\ref{g0}) for $\omega=0^+$. As $\Gamma_1 \neq \Gamma_2$ and $\gamma \neq 0$ the lead self-energy contribution affects the real (and the imaginary) part of the eigenvalues of the dot system. 
In Sec.~\ref{sec:ESOI} the lead self-energy contribution was proportional to unity and thus the real parts of the eigenenergies of the isolated dot remained unaffected. 
This renormalization by the leads has a drastic effect on the spin-orbit energy as the dashed lines in Fig.~\ref{fig:fig11} show. The overall structure remains similar to the serial symmetric case of Sec.~\ref{sec:ESOI} with a reduced maximum (compared to $2\,\alpha$ in Sec.~\ref{sec:ESOI}) at $\phi=0$ and a decrease towards larger $\phi$. The spin-orbit energy then tends towards 0 for $\phi \approx 0.45\pi$ and remains very small up until $\phi=0.5 \pi$. The picture for the interacting dot is similar (full line in Fig.~\ref{fig:fig11}) but the two-particle interaction decreases the range for which the spin-orbit energy is small compared to the $U=0$ case. In the theory plots this lead renormalization effect occurs most visibly for large $\phi$ where the spin-orbit energy is small. We refrain from comparing this effect to the experiment as it might be masked by finite temperature effects or the resolution in the bias spectroscopy. It is obvious that this effect depends strongly on the detailed geometry of the quantum dot. A more thorough analysis is needed to determine the importance of this result for experiments. 
\subsubsection*{{\bfseries{Effective $g$-factors}}}
\begin{figure}[t!]
\center{\includegraphics[clip=true,width=8.5cm]{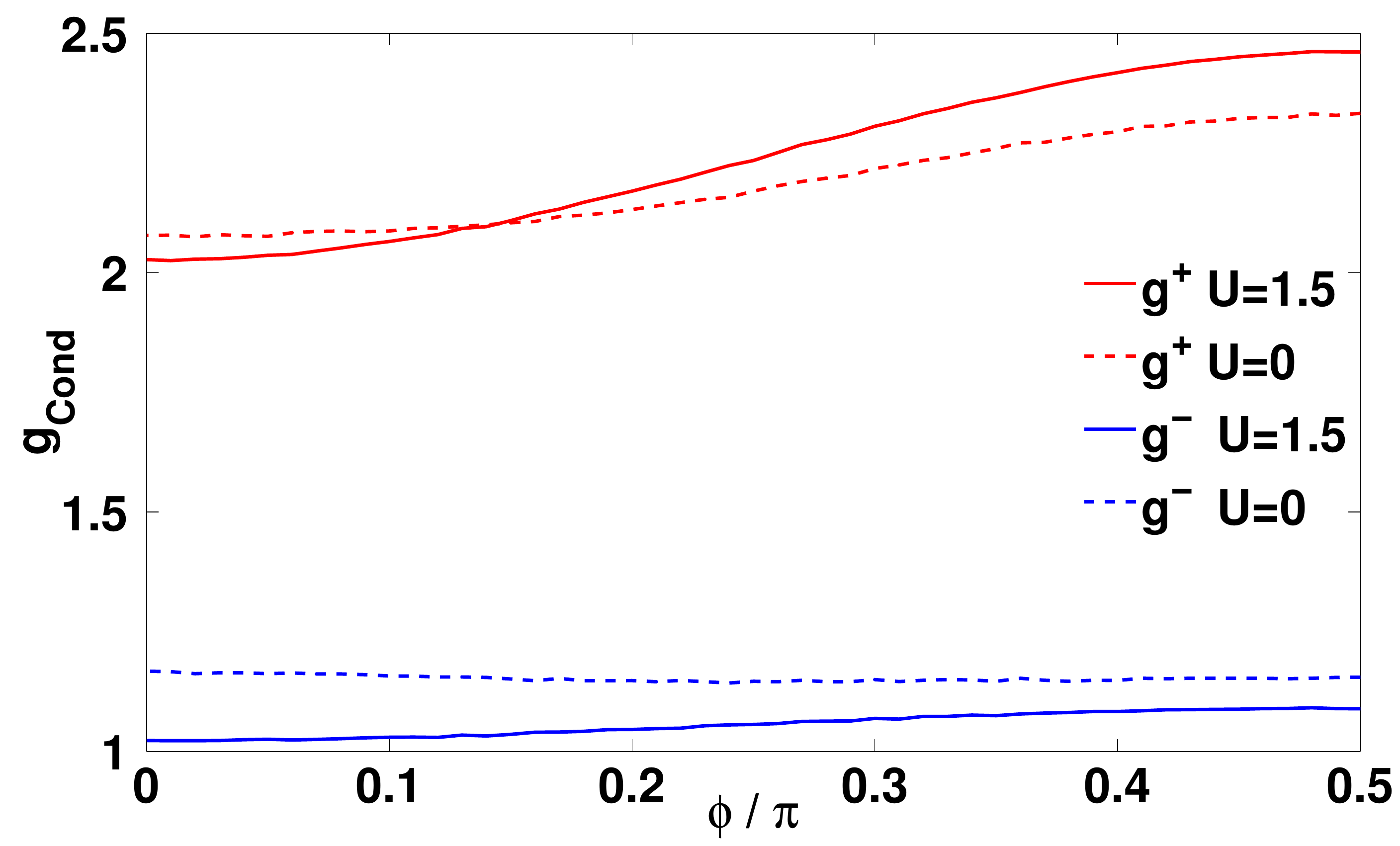}}
\caption{\label{fig:fig12} (Color online)
 Angular dependence of $g^{\pm}_{\rm Cond}$ for the parameters of Fig.~\ref{fig:fig3} (full lines). For reference the dependence of the noninteracting case is shown by the dashed lines.}
\end{figure}

For $g_{\rm Cond}$ we find several interesting features compared to the symmetric case in the previous section. The most obvious effect seen in Fig.~\ref{fig:fig12} is a different strength of the effective $g$-factor depending on which $B=0$ level degeneracy ($U=0$) or Kondo resonance ($U\neq0$) is selected for measurement ($V_{\rm G} \lessgtr 0 \mathrel{\widehat{=}} g^{\pm}$). This is already seen in the noninteracting curves depicted as dashed lines in Fig.~\ref{fig:fig12} and is only weakly affected by the interaction (full lines). An interesting effect is observed if one considers the angular dependence of the noninteracting case. Only $g^+_{\rm Cond}$ follows the general form found in Sec.~\ref{sec:g}, while $g^-_{\rm Cond}$ remains nearly constant over the whole $\phi$ range (slight deviations might be attributed to numerics). If the interaction is turned on this atypical behavior is not found and both curves follow the general form with only slightly renormalized amplitude compared to the free case. 
\begin{figure}[t!]
\center{\includegraphics[clip=true,width=8.5cm]{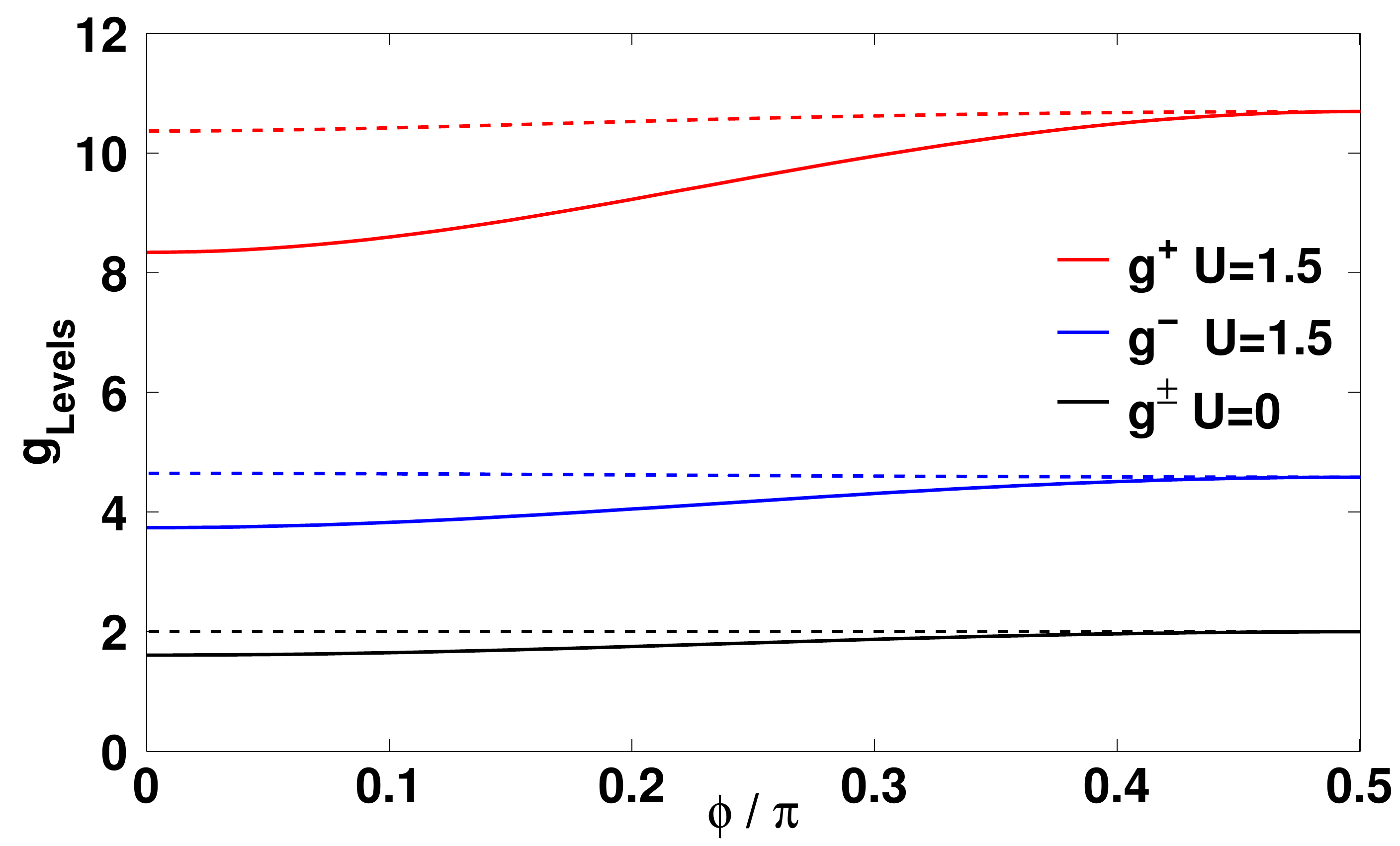}}
\caption{\label{fig:fig13} (Color online)
 Angular dependence of $g^{\pm}_{\rm Levels}$ for the parameters of Fig.~\ref{fig:fig3} (full lines). Normalization with respect to the bare angular dependence of the noninteracting case is shown by the dashed lines.}
\end{figure}

Computing $g_{\rm Levels}$ we find that the angular dependence found in Sec.~\ref{sec:g} is qualitatively preserved (full lines in Fig.~\ref{fig:fig13}). The two-particle interaction renormalization effect on the detailed functional form of the angular dependence is suppressed in comparison to the symmetric setup (dashed lines in Fig.~\ref{fig:fig13}). 
As for $g_{\rm Cond}$ a strong dependence on the asymmetry is seen if we consider both $B=0$ Kondo ridges, but only in the interacting case. As the two sets of states involved in forming the Kondo effect are coupled differently to the leads, the renormalization of parameters close to the resonance is different as well. This results in different $g_{\rm Levels}^{\pm}$ for the Kondo plateaus at positive and negative gate voltages. This asymmetry effect present in $g_{\rm Cond}$ as well as $g_{\rm Levels}$ has already been observed in experiments\cite{nilsson,csonka,kanai} where the two Kondo ridges can be attributed to different orbitals of the device. 

\section{Conclusion}\label{concl}

We studied the influence of the Coulomb interaction on the level splitting induced by the spin-orbit interaction (the spin-orbit energy), and the effective $g$-factors in multi-level quantum dots with SOI.
Furthermore, we interpret fRG results in terms of effective single-particle energy levels to obtain an intuitive physical picture for the understanding of finite-bias spectroscopy.~\cite{Fussnote} 
For a basic symmetrically coupled serial model we find that two experimentally investigated quantities are affected by the local Coulomb interaction. 
In the case of the spin-orbit energy $E_{\rm SOI}$ the effect on the overall amplitude as well as the dependence on the relative orientation between the SOI and the applied Zeeman field is very pronounced even for intermediate interaction strengths.
For the $g_{\rm Levels}$-factor calculated from the gate-voltage dependent effective level splitting - 
which mimics the experimental $g$-factor extraction from bias spectroscopy - we find sizable renormalization of its magnitude while the angular dependence is only mildly affected.
On the other hand, the $g_{\rm Cond}$-factor extracted from the Coulomb blockade peak splitting of the linear conductance appears to be almost interaction independent. The considered asymmetric parameter set shows qualitatively similar results. They nevertheless reveal a complex interplay of lead coupling, two-particle interaction and SOI.
The presented results are of importance for the understanding of transport measurements of multi-level quantum dots with SOI in presence of an external Zeeman field as reported in Refs. [\onlinecite{kanai}], [\onlinecite{deacon}], [\onlinecite{takahashi}], and [\onlinecite{nilsson}].
While the qualitative behavior of the considered quantities is consistent with the experimental data, the observed deviations and interaction-dependent amplitudes in the theoretical calculations provide new directions to investigate in future experiments. 

\section*{Acknowledgments}

We are grateful to T. Costi, S. De Franceschi, K. Grove-Rasmussen, M. Pletyukhov, 
S. Tarucha, S. Takahashi, D. Schuricht, J. Splettst\"o\ss er, and M. Wegewijs for valuable discussions.
This work was supported by the Deutsche Forschungsgemeinschaft 
(FOR 912).


\vfill\eject


\begin{thebibliography}{99}
\bibitem{note} 
	In experimental realizations the 3D structure of the device is apparent.\cite{takahashi,kanai,deacon} The present theoretical description deals with a minimal model that allows to focus on the physical mechanisms determining the interplay of Coulomb interaction effects and SOI.
\bibitem{FootnoteCouplings}
	The more general situation of complex coupling parameters is realized if a magnetic flux pierces the dot geometry. In the case considered here the couplings are related to the overlap of the dot and lead wavefunctions which can be positive or negative.
\bibitem{Fussnote} 
	In the analysis of the cotunneling experiments non-equilibrium effects are often ignored. Our calculations yield the spectrum of the dot system without non-equilibrium effects.
\end{thebibliography}
\end{document}